# Excitonic - vibronic coupled dimers: A dynamic approach

B. Esser & H. Schanz

Institut für Physik
Humboldt - Universität
Invalidenstr. 110
10099 Berlin, Germany
Tel. +49-30-2803-236, Fax +49-30-2803-238
e-mail schanz@itp02.physik.hu-berlin.de

August 4, 1994

**Abstract**

The dynamical properties of exciton transfer coupled to polarization vibrations in a two site system are investigated in detail. A fixed point analysis of the full system of Bloch - oscillator equations representing the coupled excitonic - vibronic flow is performed. For overcritical polarization a bifurcation converting the stable bonding ground state to a hyperbolic unstable state which is basic to the dynamical properties of the model is obtained. The phase space of the system is generally of a mixed type: Above bifurcation chaos develops starting from the region of the hyperbolic state and spreading with increasing energy over the Bloch sphere leaving only islands of regular dynamics. The behaviour of the polarization oscillator accordingly changes from regular to chaotic.

# 1 Introduction

The coupling between electronic and vibronic degrees of freedom in molecular and condensed media is one of the basic mechanisms influencing transfer properties of electronic excitations in these systems. The investigation of its consequences started from the polaron problem (see e. g. [1] and references therein; for exciton - phonon interaction see [2]) and continued with the study of the influence of the vibronic bath variables on the excitation transfer properties in the framework of the generalized Master equation [3] and stochastic Liouville equation approaches [4].

With the development of the theory of dynamical systems in the last decade (see e. g. [5], [6]) it has become attractive to analyze the implications of this basic mechanism employing concepts and methods of this latter field. Using a dynamic system approach one focuses on the detailed picture of the real time evolution of the relevant variables of the system. Of particular importance in this connection is the by now well established fact that systems generated by Hamiltonians with coupled degrees of freedom in many cases possess state spaces of a mixed type: regions of regular motion are embedded in a chaotic environment (see e. g. [7]). This results in completely different dynamic behaviour of the system variables depending on initial conditions and / or system parameters.

Employing such a point of view in this paper the dynamics of one of the simplest systems with coupled electronic and vibronic degrees of freedom is analyzed, namely the dissipationless transfer of an electronic excitation between two sites when each site is coupled to a vibration representing the polarization of the environment. To be definite we will consider the transfer of a Frenkel exciton between the two monomers of a molecular dimer where in each of the monomers the exciton is coupled to an intramolecular vibronic degree of freedom. The Frenkel exciton can of course be substituted by any other quasiparticle described by the same Hamiltonian. In extension of earlier approaches to the problem, where the effect of the vibronic degrees was modelled by a cubic nonlinearity (the DST-dimer originating with the paper [8] - the bifurcation of the flow line picture into a homoclinic structure and the appearance of chaos for the perturbed DST dimer were considered in [9], [10]) we will take full account of the dynamics of the vibronic variables, too. Hence our results demonstrate how the two state DST results are "embedded" in a more complete dynamic approach. For the effect of nonlinear self - trapping this problem of how DST results are embedded in a complete dynamical approach is of considerable interest. As an example we quote self - trapping at nonlinear impurity oscillators in a linear host [11].

The coupled excitonic - vibronic system is treated in a mixed quantum - classical approximation. As has become evident recently the coupling of classical and quantum degrees of freedom can lead to chaos in both the quantum and the classical subsystems even in the simplest models of this type [12]. The mixed quantum - classical description is obviously justified whenever the quantum effects in one subsystem are negligible compared to the other. Such a description, however, can also constitute an important step towards the quanti-zation of the whole system when the full quantization poses a severe challenge and the classical treatment of one part has to serve as a necessary guide. The most prominent example of an approach of this kind is the well known Born - Oppenheimer approximation. Having established a state space of both, regular and chaotic dynamics for the system in such a mixed description in a next step all the methodology of the "quantum signatures of chaos" [13] can be applied to the full quantization of the system. A system for which the correspondence between non - integrability in the mixed quantum - classical description and the properties of the fully quantized version has been already studied is e. g. the two - level atom coupled to an electromagnetic field with classical and quantum treatment of the latter (see e. g. [14]). The treatment of the excitonic - vibronic system in the quantum - classical approximation in our paper completely confirms the conclusions drawn in [12] that the simplest systems in this approximation can display chaos in both subsystems and furthermore provides an example for the field of excitonic - vibronic coupled systems.

In section 2 the model and the basic equations are formulated. Section 3 contains the fixed point analysis and the related bifurcation. In section 4 we present an analysis of the dynamic properties using Poincare sections of the coupled excitonic - vibronic flow.

## 2 Model and basic equations

We consider the full dynamics of a coupled exciton - vibronic system where the excitaton is moving between the two monomers of a molecular dimer and at each monomer the exciton is coupled to an intramolecular vibronic degree of freedom. Our model is specified by the following Hamiltonian

$$H = H_{exc} + H_v + H_{int}, \tag{1}$$

where $H_{exc}$, $H_v$ and $H_{vib}$ are the excitonic, vibronic and interaction parts, respectively. $H_{exc}$ represents a standard two site model

$$H_{exc} = \sum_n \epsilon_n c_n^* c_n + \sum_{n,m} V_{nm} c_n^* c_m \tag{2}$$

where $c_n$ is the probability amplitude of the exciton to occupy the n-th molecule and $V_{nm}$ the transfer matrix element due to standard dipole-dipole interaction, here and in what follows $n, m = 1, 2$. The vibrational part $H_v$ is taken as the sum of the energies corresponding to intramolecular vibrations at each of the monomers for which we use the harmonic approximation

$$H_v = \sum_n \frac{1}{2}(p_n^2 + \omega_n^2 q_n^2). \tag{3}$$

Here $q_n, p_n$ and $\omega_n$ are the coordinate, the canonic conjugate momentum and frequency of the intramolecular vibration of the $n$-th molecule, respectively. The interaction Hamiltonian takes into account that the exciton energy depends on the molecular configuration of the monomers as expressed by the coordinates $q_n$. Using the corresponding expansion up to first order in $q_n$ one gets

$$H_{int} = \sum_n \gamma_n q_n c_n^* c_n, . \tag{4}$$

where the $\gamma_n$ are the coupling constants. In the case when several intramolecular modes are coupled to the exciton an additional summation has to be introduced in (3), (4). Being interested in the principal effects of the excitonic - vibronic interaction we restrict the consideration to the coupling with one configurational coordinate at each molecule. In what follows we will use a mixed quantum - classical description of the dynamics of the system, describing the exciton in the quantum context and the vibronic degrees of freedom in the classical approximation.

From (1) - (4) one obtains the coupled equations of motion

$$i\dot{c}_n = (\epsilon_n - \gamma_n q_n)c_n + \sum_m V_{nm} c_m \tag{5}$$

$$\dot{q}_n = p_n \tag{6}$$

$$\dot{p}_n = -\omega^2 q_n - \gamma_n |c_n|^2 \tag{7}$$

(we set $\hbar = 1$).

One way to simplify the dynamics following from the system (5 - 7) is to assume quasistationarity in the vibrating subsystem which leads to the DST approximation mentioned in the introduction. Formally this is achieved by assuming for the vibronic part $\dot{q}_n \cong 0$, $\dot{p}_n \cong 0$ which results in $q_n = -(\gamma_n/\omega_n^2)|c_n|^2$ (the vibronic coordinate $q_n$ is "slaved" by the excitonic occupation $|c_n|^2$). Inserting the latter expression into (5) one obtains

$$i\dot{c}_n = (\varepsilon_n - \chi_n |c_n|^2)c_n + \sum_m V_{nm} c_m \tag{8}$$

where $\chi_n$ will be given in (16). Eq. (8) is called the discrete self trapping (DST) equation. The transfer dynamics resulting from it were intensively analyzed for different parameter sets and numbers of molecular units $N, n = 1,..N$ (see e. g. [16]). An equation of the type of (8) appears already in the early papers on the polaron problem as a limiting case [1]. One of the aims of this paper is to relate the dynamics following from eq. (8) for the dimer, i. e. $N = 2$, which was studied in [9], to the dynamics of the full system (5) - (7). We note that an intuitive statement concerning the validity of the DST approximation is usually given in the form that it corresponds to the case of fast polarization / slow transfer and that then the vibronic subsystem is always in its ground state. One way to achieve this is the introduction of a fast relaxation into the vibronic subsystem. This, however, is beyond the scope of this paper where we remain on the dissipationless Hamiltonian level of description. For the role of dissipation in connection with the DST approximation we refer to the recent paper [15] and the references therein.

Introducing the density matrix elements

$$\rho_{mn} = c_n^* c_m \tag{9}$$

and passing to the Bloch variables

$$
\begin{align}
x &:= \rho_{12} + \rho_{21} \\
y &:= i(\rho_{21} - \rho_{12}) \\
z &:= \rho_{22} - \rho_{11}
\end{align}
\tag{10}
$$

one obtains from (5) to (7) for the excitonic subsystem the Bloch equations

$$
\begin{align}
\dot{x} &= -\Delta\epsilon(t)y \\
\dot{y} &= z + \Delta\epsilon(t)x \\
\dot{z} &= -y
\end{align}
\tag{11}
$$

where $\Delta\epsilon(t)$ is given by

$$\Delta\epsilon(t) := \frac{\gamma_2 q_2(t) - \gamma_1 q_1(t) + \epsilon_2 - \epsilon_1}{2V} \tag{12}$$

In (11) we have introduced a dimensionless time by setting $t \to 2Vt$. Obviously, the dimensionless quantity $\Delta\epsilon(t)$ controlling the coupling between the excitonic and vibronic subsystems is the dynamic generalization of the difference between the excitonic energies at each of the molecules caused by the interaction with the intramolecular oscillators. For a two site system (11) is equivalent to (5) but has the advantage of excluding the global phase.

The coupled system of Bloch - oscillator equations (6), (7), (11) has two integrals of the motion: the energy $H = E$ and the radius of the Bloch sphere $x^2 + y^2 + z^2 = 1$. Correspondingly, the number of independent variables of the system is reduced by two resulting in a total of five. A further reduction of the number of variables can be achieved by observing that only an effective oscillator, which is a linear combination of $q_1$ and $q_2$, enters the expression for $\Delta\epsilon(t)$ : After the introduction of dimensionless oscillator coordinates $Q_n := \frac{\omega_n^2}{\gamma_n}q_n$, momenta $P_n := \frac{\omega_n}{\gamma_n}p_n$ and a dimensionless energy difference

$$a := \frac{\epsilon_2 - \epsilon_1}{2V} \tag{13}$$

the quantity $\Delta\epsilon(t)$ is representable in the form

$$\Delta\epsilon(t) = Q + a \tag{14}$$

where

$$Q := \chi_2 Q_2 - \chi_1 Q_1 \tag{15}$$

and

$$\chi_n := \gamma_n^2/\omega_n^2. \tag{16}$$

were introduced. By using the oscillator equations (6) and (7) it is easily verified that the equations for $Q$ and the corresponding momentum

$$P := \chi_2 P_2 - \chi_1 P_1 \tag{17}$$

close, if the additional condition $\omega_1 = \omega_2$ is satisfied. In what follows we will assume that this condition is fulfilled, i. e. the oscillators have identical frequencies. We stress that the dimer can still be "electronically disordered" when it has different excitonic energies $\epsilon_n$ and coupling constants $\gamma_n$. Using the condition $\omega_1 = \omega_2$ one obtains the equations of motion for the effective oscillator

$$\begin{aligned} \dot{Q} &= rP \\ \dot{P} &= -rQ - r(pz + q) \end{aligned} \qquad (18)$$

where the dimensionless paramters

$$\begin{aligned} p &:= \frac{\chi_2 + \chi_1}{4V} \\ q &:= \frac{\chi_2 - \chi_1}{4V} \\ r &:= \omega/2V \end{aligned} \qquad (19)$$

were introduced. $r$ is a parameter translating between excitonic and vibronic time scales. Assuming quasistationarity for the effective oscillator and inserting the resulting equation $Q = -(pz + q)$ into (14) one obtains the Bloch subsystem (11) in DST approximation as in [9]. The five equations (11) and (18) coupled via (14) still have two integrals of the motion: besides the radius of the Bloch sphere the dimensionless energy $e := E/V$ of the considered subsystem is conserved. It can be expressed as the sum of the contribution from an effective oscillator, whose equilibrium position depends on the current state of the excitonic subsystem, and a second exclusively excitonic term:

$$e = -x - (q - a)z - \frac{pz^2}{2} + \frac{1}{2p}\left[(Q + q + pz)^2 + P^2\right] \qquad (20)$$

We note, that the condition imposed on $Q$ and $P$ to yield the DST approximation implies that the effective oscillator has zero energy. For the DST dimer it is therefore straightforward to obtain from (20) the analytical solution for the orbits on the Bloch sphere. For the full dynamics which we consider here this can be generalized to an energy dependent restriction for the accessible part of the Bloch sphere: Since the energy of the effective oscillator is non-negative, the total energy is an upper bound to the excitonic term. This results in

$$x \geq -\frac{p}{2}z^2 - (q - a)z - e, \qquad (21)$$

where the equal sign just describes the DST orbits.

## 3 Fixed point analysis and bifurcation

### 3.1 General equations

We perform the fixed point analysis including the case of a nonsymmetric dimer ($\epsilon_1 \neq \epsilon_2, \gamma_1 \neq \gamma_2$). Setting in the equations of motion the time derivatives of $z$,

$Q$ and $P$ equal to zero, we find for the stationary states

$$
\begin{aligned}
y_s &= 0 \\
Q_s &= -(pz_s + q) \\
P_s &= 0
\end{aligned}
\qquad (22)
$$

The values for $x_s$ and $z_s$ are determined by $\dot{y} = 0$. Using (22) one obtains

$$(a - q)x_s + z_s - px_s z_s = 0 \qquad (23)$$

The restriction of the excitonic variables to the Bloch sphere yields the additional condition

$$x_s^2 + z_s^2 = 1. \qquad (24)$$

In order to determine the character of the fixed points we need the linearization of the dynamics for small deviations from them. The eigenvalues of the corresponding stability matrix turn out to obey the characteristic equation

$$\lambda^4 + (x_s^{-2} + r^2)\lambda^2 + r^2(x_s^{-2} - px_s) = 0. \qquad (25)$$

with the location of the fixed point still general. According to the Hamiltonian character of the problem it is biquadratic and the sum of all stability exponents $\lambda$ is zero. They come in pairs

$$\lambda^2 = -\frac{x_s^{-2} + r^2}{2} \pm \frac{1}{2}\sqrt{(x_s^{-2} - r^2)^2 + 4pr^2 x_s}. \qquad (26)$$

It is appropriate to subdivide all stationary points according to whether they are located in the bonding ($x_s > 0$, containing the bonding state $x_s = +1$) or antibonding region ($x_s < 0$, containing the antibonding state $x_s = -1$). There is no transition between the two groups since $x_s = 0$ is excluded by (23).

Let's consider the bonding region first. From (26) is evident that in this region $\lambda^2$ is always real for both pairs of stability exponents and negative for at least one of them. The sign of the other one can be positive or negative. In the latter case all stability exponents are purely imaginary and the fixed point is stable elliptic. The condition for this to hold is

$$p < x_s^{-3}. \qquad (27)$$

If it is not fulfilled the fixed point has a hyperbolic manifold and is unstable.

In the antibonding region $\lambda^2$ itself can be complex if

$$|r^2 - x_s^{-2}| < 2r\sqrt{p|x_s|}. \qquad (28)$$

In this case the antibonding fixed point will be unstable.

From (22) it is clear that the effective oscillator in (20) has zero energy at every fixed point. Consequently, the fixed points of the full problem are identical with those from the DST problem [9] and we can use the DST orbits from (21) to show their location on the Bloch sphere. We stress, that the equivalence to the DST approximation does not hold for the stability: The applicability of a time scale separation leading to the DST dimer depends on the value of the parameter $r$ which does enter neither the location of fixed points (23) nor the stability condition for the bonding region (27) but the stability in the antibonding region (28). Consequently, the fixed point in the antibonding region was found to be always stable in [9] whereas it can be unstable in our more general approach.

## 3.2 Symmetric dimer

Finding the solutions of (23) it is convenient to consider the case of a symmetric dimer ($\epsilon_1 = \epsilon_2$, $\chi_1 = \chi_2$, i. e. $a = q = 0$) first. Then the stationarity condition (23) depends on the parameter $p$ only and reduces to the form

$$(1 - px_s)z_s = 0 \qquad (29)$$

We note that (29) also includes the situation when there is an exact compensation of the energy mismatch $a$ and the polarization difference $q$. Considering the solutions of (29) one has to take account of (24) restricting $x_s$ and $z_s$ to values $x_s \leq 1$ and $z_s \leq 1$. So one finds the following solutions in dependence on the value of the parameter $p$:

- **Bonding region** ($x_s > 0$):

  - **Case A:** $0 \leq p < 1$ In this case the only solution of (29) is $z_s = 0$ and one finds from (24) the stationary point g with $x_s = +1$. The point g is the ground state corresponding to a symmetric combination of excitonic amplitudes $c_1 = c_2 = 1/\sqrt{2}$. The stability condition (27) is fulfilled.

  - **Case B:** $p = 1$ In this special situation we find two additional fixed points from the solution of (29) which are degenerate with the point g.

  - **Case C:** $p > 1$ There are two solutions of (29): $z_s = 0$ and $x_s = 1/p$. From (24) we get for the first case $x_s = \pm 1$ and for the second case $z_s = \pm\sqrt{1 - (1/p)^2} =: \pm z_0$. Correspondingly, we obtain three stationary points

    g1: $x_s = 1/p; z^+ = +z_0$
    g2: $x_s = 1/p; z^- = -z_0$
    h: $x_s = +1; z_s = 0$

    The points g1 and g2 are stable, whereas h is unstable hyperbolic.

The comparison of the structure of stationary points in the cases A and C reveals that the parameter $p$ governs a pitchfork bifurcation: The ground state g below the bifurcation ($p < 1$) splits into two degenerate ground states g1 and g2 above bifurcation ($p > 1$). At the former ground state a hyperbolic point h appears. This is also evident from fig. 1. The DST trajectories shown there can be interpreted in the full problem as isoenergy level lines within the subspace of zero oscillator energy. The physical reason for the bifurcation is that for $p > 1$ the polarization is strong enough to hold the excitation preferentially at one of the molecules of the dimer. This corresponds in fig. 1 (b) to closed lines around the degenerate ground states. The regions on the Bloch sphere encircled by them can never be left by the exciton since the accessible part of the Bloch sphere as given by (21) will shrink further if the vibronic subsystem is not in the state of minimal energy. In this way we can conclude from fig. 1 (b) that above bifurcation there exist trajectories in the higher dimensional phase space of the full excitonic - vibronic problem where the exciton is self trapped, i. e. with a time average of $z$ different from zero.

- **Antibonding region** ($x_s < 0$): In this region the only solution of (23) is $z_s = 0$, $x_s = 1$. This is the excited stationary state e and corresponds to an antisymmetric combination of excitonic amplitudes $c_1 = -c_2 = 1/\sqrt{2}$. According to (28) this point is stable for $\frac{|r^2|-1}{r} > 2\sqrt{p}$ which will hold for the limiting cases $r \ll 1$ and $r \gg 1$. If, however, the phonon and transfer frequencies are close to each other ($r \sim 1$) the stability is lost.

## 3.3 Asymmetric dimer

We now turn to the solution of eq. (23) for the general case of an asymmetric dimer. Without loss of generality the difference $q - a$ can be considered one parameter. Therefore we will absorb the parameter $a$ into a new defined $q := q - a$. Inserting $x_s = \pm\sqrt{1 - z_s^2}$ into (23) one obtains the following cases:

- **Bonding region** ($x_s > 0$):

$$q + pz_s = \frac{z_s}{\sqrt{1-z_s^2}} \tag{30}$$

In this case there are one and three solutions possible. In fig. 2 we present plots corresponding to these cases. In between there is the special situation when the straight line on the l.h.s. of (30) is tangent to the r.h.s. The geometrical condition of tangency in the intermediate situation can be used to derive the bifurcation diagramm fig. 3 in the $q, p$ parameter plane. One obtains two lines $\pm q(p)$ separating the region with a single solution for (30) from the one with three solutions. The lines are given by

$$q(p) = (p - p^{1/3})(1 - p^{-2/3})^{1/2}, \qquad p \geq 1 \tag{31}$$

Starting from the symmetric dimer and using (27) one can show that in the case of three fixed points two are always stable and one is unstable hyperbolic. If there is just one fixed point, it must be stable elliptic. Thus the bifurcation diagramm generalizes the bifurcation of the ground state as expressed by the three cases A, B and C for the symmetric dimer to the asymmetric case. In particular, for $p > 1$ and $q$ within the area restricted by the two lines in fig. 3 the former ground state has split into two new elliptic and one hyperbolic state. In the asymmetric case these elliptic states have, however, different energies, i. e. the ground state degeneracy is destroyed. We note the interesting effect that with increasing asymmetry, i. e. increasing value of $|q|$, the region of three fixed points is left and consequently the bifurcation is inverted. This is also evident from fig. 4 where the positions of $z_s$ and $x_s$ for a fixed $p > 1$ are presented in dependence on $q$. We note that for a large linear asymmetry realized by $|a| \gg |q| \sim p$ the solutions of (23) approach the values for the linear asymmetric dimer ($x_s \to \pm(1 + a^2)^{-1/2} \to 0$, $z_s \to \pm a(1 + a^2)^{-1/2} \to \pm 1$).

- **Antibonding region** ($x_s < 0$):

$$q + pz_s = -\frac{z_s}{\sqrt{1 - z_s^2}} \tag{32}$$

This equation has only one solution corresponding to the excited stationary state. It is stable or unstable according to whether (28) is fulfilled or not.

The location of the fixed points and the bifurcation for the asymmetric case is illustrated in fig. 5 for one special value of the asymmetry parameter $q$.

As already stated above, the stationarity condition for the full set of variables contains DST stationarity as a special case. Comparing the bifurcation with the one obtained in the DST approximation of the transfer problem (see [9]) one therefore finds that it is of the same pitchfork type represented by a ground state splitting and the appearance of a hyperbolic point.

## 4 Analysis of the phase space flow

### 4.1 General remarks

Before turning to a detailed investigation of the dynamical regimes of our model we mention some general points concerning the coupled excitonic - vibronic flow. First of all we note that this flow is realized on a foliated phase space consisting of the surface of the Bloch sphere and the phase plane representing the dynamic states of the exciton and the oscillator, respectively. The flow of the excitonic and vibronic variables is restricted to each of the subspaces which constitute the "leaves" of the foliated space. On each of these two dimensional subspaces the dynamics of a one degree of freedom subsystem is realized. Taking account of the energy conservation three of the four dynamic variables of the coupled systems are independent. Hence by fixing one of the three independent variables a Poincare section on a two dimensional submanifold is determined. Depending on the choice of the independent variables and which of them is fixed we have Poincare sections realized on (i) the surface of the Bloch sphere, (ii) the phase plane of the oscillator and (iii) mixed manifolds.

In the next subsection we demonstrate the characteristic dynamic regimes connected with the presence of the bifurcation and in particular the appearance of a hyperbolic point at the former ground state. These effects are displayed by the symmetric as well as by the asymmetric dimer and hence are presented for the simpler symmetric case. For the same reason we demonstrate the effect of the resonance in the simplest case of a 1-1 resonance for the symmetric dimer.

Considering the asymmetric dimer we found flow pictures which may be thought as a continous deformation of the symmetric case. They do not contain qualitatively new results and are therefore not displayed. However, there are some special effects connected with the two dimensional character of the bifurcation diagram which will be discussed in the last subsection.

## 4.2 Symmetric dimer

**Below bifurcation**

In fig. 6 (a) and (b) Poincare sections on the surface of the Bloch sphere are presented for two different energies and below the bifurcation ($p = 0.5 < 1$). They correspond to the left turning point of the effective oscillator ($P = 0$, $dP/dt > 0$). Poincare sections at the right turning point contain the same information as one can see from a symmetry property: From every trajectory in the symmetric dimer another trajectory can be found by replacing $z$ and $Q$ with $-z$ and $-Q$, respectively, since the total energy (20) is invariant under this transformation. From the Poincare sections one finds regular dynamics for both, low and high energy. For low energy the elliptic structure is centered in the region of the bonding ground state ($x = 1, \phi = 0$) whereas for high energy an analogous elliptic structure in the the region of the antibonding excited state ($x = -1, \phi = \pi$) is obtained. The oscillator also displays regular dynamics as is shown in the Poincare section of fig. 7.

**Above bifurcation**

In figs. 8 (a) - (d) a typical set of Poincare sections corresponding as in fig. 6 to the left turning point of the oscillator but for a polarization strength above bifurcation ($p = 2.0 > 1$) is shown. For low energy one finds regular dynamics in the region of the new ground states ($x_s = 1/p, z = \pm z_0$) as in fig. 6a. These regular trajectories are "trapped" solutions of the excitonic subsystem in which the exciton is preferentially at one of the sites of the dimer. Their time average of $z$ is $\overline{z} \sim \mp z_0$ for the exciton being trapped at the first or second site, respectively. Correspondingly, the polarization oscillator performs small oscillations around the equilibrium positions $Q = \pm pz_0$.

For intermediate energies involving the hyperbolic point, i. e. the dimensionless energy $e = -1$ which can be used as a reference energy separating the low and high energy regions, one still obtains regular behaviour. From fig. 8 (b) the structure of the hyperbolic point is evident. In fig. 8 (c) and (d) the spreading of chaotic dynamics over the Bloch sphere with increasing energy is shown. Beginning at the hyperblic point more and more trajectories become detrapped, i. e. the time average for $z$ tends to zero with increasing energy. Chaos starts in the vicinity of the hyperbolic point ($x = 1$, $z = \phi = 0$, $e = -1$) and then spreads over the Bloch sphere leaving only regular islands in the region of antibonding states. We observe that for chaotic trajectories the energy of the polarization oscillator is relatively high whereas for the regular orbits around the antibonding states the energy of the excitonic subsystem is at its highest level and the oscillator does not get enough energy to destroy the regular structure of the motion. This point of view is confirmed by the Poincare section in oscillator variables which is displayed in fig. 9: For small energies of the oscillator the dynamics is regular, whereas in the outer region of the oscillator phase plane where its energy is high the trajectories are chaotic.

**Resonance situation**

Another important feature displayed by the full Bloch - oscillator system is the possibility of resonance between the excitonic and vibronic subsystems. In the DST approximation this effect is absent due to the quasistationarity of the vibronic subsystem. In fig. 10 Poincare sections corresponding to such a resonance are displayed ($r = 1.0$, i. e. the transfer frequency equals the vibronic frequency $2V = \omega$). It should be noted that the coupling strength $p = 0.5 < 1$ is well below the bifurcation value and so there is no hyperbolic point in the bonding region. The antibonding fixed point, however, is unstable due to the resonance and therefore the dynamics becomes irregular as the energy enters the antibonding region.

## 4.3 Asymmetric dimer

The two dimensional character of the bifurcation in the case of an asymmetric dimer and the dynamic behaviour of the system associated with it is illustrated in the fig. 11. The coupling strength ($p = 2$) is chosen at a value where the symmetric dimer would be above the bifurcaion. The asymmetry parameter ($q = 0.5$) is sufficiently large that the $(p, q)$ pair is outside the bifurcation region. Correspondingly the dynamics of the system is regular for intermediate energies as in fig. 11a. The effect of the asymmetry is, however, supressed by a high oscillator energy as it can be seen in fig. 11b where the dynamics for the same $(p, q)$ parameter set but at much higher total energy is displayed. There is no important difference compared to the symmetric dimer above bifurcation from fig. 8c, d. This shows that in the high energy region the dynamic part $Q$ is much larger than the static difference $a$ contributing to the shift in the site energies (14) and the asymmetry of the coupling strengths $q$ in (18). Hence the system is effectively symmetric and behaves irregular because of a coupling strength $p > 1$.

## 5 Conclusions

1. The simplest model of electronic excitation transfer coupled to vibronic degrees of freedom, namely the excitation transfer between two sites coupled to a polarization oscillator possesses a phase space with mixed properties: The system displays chaotic or regular behaviour depending on the initial conditions on the Bloch sphere and the coupling strength to the polarization oscillator. Basic to the properties of the system is the bifurcation converting the former ground state in the bonding region of the Bloch sphere for a coupling strength above a critical value into an unstable hyperbolic point from which chaotic dynamic starts to develop and spreads over the Bloch sphere with increasing total energy of the system. The region of antibonding states on the Bloch sphere is the region of highest energy in the electronic subsystem where compared to the bonding region for a fixed total energy the energy of the oscillator subsystem is lowest and the regular structures persist destruction until the total energy is well above that of the antibonding states. Besides high energy and overcritical polarization there is a second mechanism for irregular dynamics: resonance between the electronic and vibronic subsystems.

2. The character of the bifurcation is not changed when asymmetry effects in the coupling strength and electronic site energies are introduced. The bifurcation diagram becomes, however, two dimensional containing an additional asymmetry parameter. The asymmetry parameter can vanish when the difference of the coupling strengths is exactly compensated by the energy mismatch of the site energies. There is also a dynamic supression of asymmetry possible when the changes in the site energies produced by the irregular motion of the oscillator in the chaotic region are much larger than the differences in the coupling strength and / or site energies.

3. In the DST approximation the polarization oscillator is assumed to be always in its ground state. The type of the bifurcation of the fixed points is the same for the full system and the DST case. In the latter case, however, the bifurcation is realized on the integrable flow of two independent variables on the Bloch sphere. The hyperbolic point reduces to the crossing of the separatrices in a homoclinic structure on the Bloch sphere. The perturbation of this homoclinic structure by a non - integrable perturbation is a well known route to chaos. Considering the dynamics of the polarization oscillator as such a perturbation the DST solutions can be used as a reference to understand the onset of chaotic behaviour in the hyperbolic point / separatrix region.

# 6 Acknowledgement

Support from the Deutsche Forschungsgemeinschaft (DFG) is gratefully acknowledged.

# References

[1] J. Appel: Solid State Physics **21**, 193 (1968)

[2] A. S. Davydov: "Theory of Molecular Excitons", Plenum, New York (1979)

[3] V. M. Kenkre: "The Master Equation Approach" in "Molecular Crystals and Aggregats", Springer Tracts in Modern Physics **Vol. 94**, Springer - Verlag (1982)

[4] P. Reineker: "Stochastic Liouville Equation Approach" in see [3]

[5] J. Guckenheimer, Ph. Holmes: "Nonlinear Oscillations, dynamical systems and bifurcation of vector fields" in Applied Mathematical Sciences, Vol. **42**, Springer Verlag (1986)

[6] S. Wiggins: "Chaotic Transport in Dynamical Systems", Springer Verlag, Berlin (1992)

[7] M. Tabor: "Chaos and Integrability in Nonlinear Dynamics", Wiley and Sons (1989)

[8] V. M. Kenkre, D. K. Campbell: Phys. Rev. **B34**, 4595 (1986)

[9] B. Esser, D. Hennig: Z. Phys. B (Condensed Matter) **83**, 295 (1991)

[10] D. Hennig, B. Esser: Phys. Rev. **A46**, 4569 (1992)

[11] D. Chen, M. I. Molina, G. P. Tsironis: J. Phys.: Condensed Matter **5**, 8689 (1993)

[12] R. Blümel, B. Esser: Phys. Rev. Lett. **72**, 3658 (1994)

[13] F. Haake: "Quantum Signatures of Chaos", Springer Verlag, Berlin (1991)

[14] R. Graham, M. Höhnerbach: Phys. Lett. **A101**, 61 (1984)

[15] D. Vitali, P. Allegrini, P. Grigolini: Chem. Phys. **180**, 297 (1994)

[16] J. C. Eilbeck, P. S. Lomdahl, A. C. Scott: Physica, **D16**, 318 (1985)